\documentclass[%
 aip,
cp,  
 amsmath,amssymb,
 reprint,%
]{revtex4-2}

\usepackage{graphicx}
\usepackage{dcolumn}
\usepackage{bm}

\usepackage[utf8]{inputenc}
\usepackage[T1]{fontenc}
\usepackage{mathptmx} 

\begin{document}

\title{The Phenomenology of Sphaleron\\ in Modified Mirror Model}

\author{Apriadi Salim Adam} 
\email[Corresponding author: ]{apriadi.salim.adam@brin.go.id}
\affiliation{
  Research Center for Quantum Physics, National Research and Innovation Agency (BRIN), South Tangerang 15314, Indonesia
}

\author{Firdaus}%
\email{firdaus@fkip.untan.ac.id}
\affiliation{%
Department of Physics Education, Faculty of Education, Tanjungpura University, 78124 Pontianak, Indonesia
}%

\author{Mirza Satriawan}
\email{mirza@ugm.ac.id}
\affiliation{Department of Physics, Universitas Gadjah Mada, Bulaksumur BLS 21 Yogyakarta 55281, Indonesia}

\date{\today} 

\begin{abstract}
We investigate sphaleron solutions of the field equations in the modified mirror model. This model is based on SU(3)$_1$ $\otimes$ SU(3)$_2$ $\otimes$ SU(2)$_L$ $\otimes$ SU(2)$_R$ $\otimes$ U(1)$_{Y}$ $\otimes$ U(1)$_{X}$ gauge group. Different from the usual Weinberg-Salam theory, we will have two types of fields, the ordinary (standard model) and its mirror partners. In this paper, we will focus only on the case when the doublet scalar of the mirror sector has a very large vacuum expectation value (VEV) compared to the VEV of the SM Higgs. We study two scenarios related to the sphaleron solutions in this model, depending on the value of the coupling $\alpha$. For the case of $\alpha=0$, each sector has its sphaleron and the sphaleron energy depends on the VEV of scalars for each sector. In the case $\alpha \neq 0$, we find that the sphaleron energy in the SM sector can be either increased or decreased depending on the relative sign of the coupling $\alpha$.
\end{abstract}

\maketitle

\section{Introduction}
Among asymmetric dark matter model, mirror model is the most promising one due to its simple concept with relatively small number of additional parameters (see \cite{foot1997neutrino,foot2000unbroken} and \cite{foot2014mirror} for a review on the mirror model and \cite{Davoudiasl:2012uw,Petraki:2013wwa,Zurek:2013wia,Lonsdale:2018xwd} for review on asymmetric dark matter).  As an asymmetric dark matter model, the dark matter in the mirror model is the leftover particles due to particle-antiparticle asymmetry in the mirror sector.  Thus, the mirror model assumes that there is some baryogenesis or leptogenesis process in the mirror sector similar to the one in the ordinary or standard model (SM) sector.   In the original mirror model, the mirror symmetry is not broken and the two sectors are the same, i.e. the mirror particles will have the same mass spectra as in the SM sector.  This is due to the same vacuum expectation value (VEV) for the SU(2) scalar doublet, i.e. the Higgs and mirror Higgs.  The two sectors will experience the same leptogenesis and also have the same B-violating process that converts the lepton number produced during leptogenesis into baryon number.  In such a model, the relative abundance of the dark matter $\Omega_{DM}$ should be the same as the baryonic one $\Omega_{B}$, but in reality, it is known that $\Omega_{DM} \simeq 5 \Omega_{B}$ \cite{Planck:2018vyg}. This, together with the constraint on the number of relativistic degree of freedom in the early cosmology (from the big bang nucleosynthesis, large-scale formation, cosmic microwave background radiation, etc.) suggest that if the mirror model is the correct candidate for the dark matter model, the mirror symmetry has to be broken at some high energy scale.  

Since any leptogenesis should take place after the cosmological inflation (otherwise the produced lepton number will be diluted), any B-violating process that converts the lepton number into baryon number should take place after the leptogenesis and before the process ends.  It is known that the sphaleron solution that may occur before the electroweak phase transition can provide such a B-violating process that preserves $B-L$.   Therefore to make $\Omega_{DM}$ not equal to the $\Omega_B$, during the occurrence of the sphaleron process the mirror symmetry should already be broken.  We refer to a model with broken mirror symmetry as the modified mirror model (MMM).  The mirror symmetry breaking could take place before the cosmological inflation, or after the cosmological inflation, but it should take place before the occurrence of the sphaleron process in the mirror model.  In this paper, we present our investigation on the sphaleron solution or the saddle point solution in the modified mirror model.  We only consider the case where the doublet scalar of the mirror sector has a very large VEV compared to the VEV of the SM Higgs. 

A typical MMM usually has particle content as in table \ref{tabelmassa1} with a Lagrangian that is invariant under the mirror gauge group SU(3)$_1$ $\otimes$ SU(3)$_2$ $\otimes$ SU(2)$_L$ $\otimes$ SU(2)$_R$ $\otimes$ U(1)$_{Y}$ $\otimes$ U(1)$_{X}$ and the Z$_2$ mirror symmetry. In this paper, we assume that the mirror symmetry has broken at some high energy scale and the effect of the mirror breaking causes soft mirror-breaking terms (the VEV of the SM Higgs is different from the VEV of the mirror Higgs). 
\begin{table}
\caption{\label{tabelmassa1}Irreducible representation (Irreps) and quantum numbers assignment for the scalar and spinor particles with respect to the mirror gauge group, SU(3)$_1$$\otimes$SU(3)$_2$$\otimes$SU(2)$_L$$\otimes$SU(2)$_R$$\otimes$U(1)$_{Y}$$\otimes$U(1)$_{X}$. Note that the last row is the scalar.}
\begin{ruledtabular}
\begin{tabular}{ccccc}
SM-particles & Irreps & & m-particles & Irreps\\
\hline
 & & & & \\
$l_L \equiv \left( \begin{array}{l}
\nu \\
e
\end{array}\right)$ & (\textbf{1},\textbf{1},\textbf{2},\textbf{1},-1,0) & &
$L_R \equiv \left( \begin{array}{l}
N \\
E
\end{array}\right)$ &(\textbf{1},\textbf{1},\textbf{1},\textbf{2},0,-1) \\ 
$\nu_R$ &(\textbf{1},\textbf{1},\textbf{1},\textbf{1},0,0)& & $N_L$ & (\textbf{1},\textbf{1},\textbf{1},\textbf{1},0,0) \\ 
 $e_R$ & (\textbf{1},\textbf{1},\textbf{1},\textbf{1},-2,0)& &$E_L$ & (\textbf{1},\textbf{1},\textbf{1},\textbf{1},0,-2) \\ 
$q_L \equiv \left( \begin{array}{c}
u \\
d
\end{array}\right)$ &(\textbf{3},\textbf{1},\textbf{2},\textbf{1},$\frac{1}{3}$,0) & &
$Q_R = \left( \begin{array}{c}
U \\
D
\end{array}\right)$ &(\textbf{1},\textbf{3},\textbf{1},\textbf{2},0,$\frac{1}{3}$)\\
$u_R$ & (\textbf{3},\textbf{1},\textbf{1},\textbf{1},$\frac{4}{3}$,0)& & $U_L$ & (\textbf{1},\textbf{3},\textbf{1},\textbf{1},0,$\frac{4}{3}$)\\ 
$d_R$ & (\textbf{3},\textbf{1},\textbf{1},\textbf{1},$\frac{-2}{3}$,0)& &
  $D_L$ & (\textbf{1},\textbf{3},\textbf{1},\textbf{1},0,$\frac{-2}{3}$) \\
			& & & & \\
			\hline 
		& & & & \\
	$\chi_L = \left( \begin{array}{c}
\chi_\nu \\
\chi_e
\end{array}\right)$ &(\textbf{1},\textbf{1},\textbf{2},\textbf{1},-1,0)& &
$\chi_R = \left( \begin{array}{c}
\chi_N \\
\chi_E
\end{array}\right)$ &(\textbf{1},\textbf{1},\textbf{1},\textbf{2},0,-1)
\end{tabular}
\end{ruledtabular}
\end{table}

\section{Sphaleron in Modified Mirror Model}
To investigate the sphaleron process in the MMM, we start with the following Lagrangian for the gauge-Higgs sector,
\begin{eqnarray}
  \mathcal{L}_{\text{MMM}} & = & - \frac{1}{4} F_{L \mu \rho}^a F_L^{a \mu \rho}
  - \frac{1}{4} F_{R \mu \rho}^a F_R^{a \mu \rho} - \frac{1}{4} f_{\mu \rho L}
  f^{\mu \rho L} - \frac{1}{4} f_{\mu \rho R} f^{\mu \rho R} \nonumber\\
  &  & + (D_{\mu} \chi_L)^{\dag} (D^{\mu} \chi_L) + (D_{\mu} \chi_R)^{\dag}
  (D^{\mu} \chi_R) - V (\chi_L, \chi_R),  \label{lag0}
\end{eqnarray}
where $F_{L (R) \mu \rho}^a$ are the field strength of the SU$(2)_{L (R)}$, $f_{\mu \rho L (R)}$ is the field strength of the U$_{Y (X)}$ and $D_{\mu} \chi_{L(R)}$ is the covariant derivative for the left (right) doublet scalars, respectively. In the above equation, the scalar potential $V (\chi_L, \chi_R)$ is invariant under the gauge and $Z_2$-mirror transformations and it is given by,
\begin{eqnarray}
  V (\chi_L, \chi_R) & = & \lambda \left[ \left( | \chi_L |^2 - \frac{1}{2}
  v^2_L \right)^2 + \left( | \chi_R |^2 - \frac{1}{2} v^2_R \right)^2 \right]
  + \alpha | \chi_L |^2 | \chi_R |^2 
  .  \label{pot1}
\end{eqnarray}
If all of the couplings are non-negative, then $V (\chi_L, \chi_R) \geqslant 0$. The scalar fields can gain their vacuum expectation values as follows,
\begin{eqnarray}
  \langle \chi_L \rangle   & = & \frac{1}{\sqrt{2}}
  \left(\begin{array}{c}
    v_L\\
    0
  \end{array}\right), \\
  \langle \chi_R \rangle & = & \frac{1}{\sqrt{2}} \left(\begin{array}{c}
    v_R\\
    0
  \end{array}\right). 
\end{eqnarray}

Next, we will consider the static classical (sphaleron) solutions for the fields in the MMM. For this purpose, both the fermions and the time component of the gauge fields are set to zero. Following the same procedure in \cite{Klinkhamer:1984di,Akiba:1988ay,Kastening:1991nw}, we shall set $\theta_{W_{L (R)}} = 0$ so that the U$_{Y (X)}$ gauge fields are decoupled from the scalars and may be consistently set to zero. In this setup, the energy functional is written as follows,
\begin{eqnarray}
  E_{\text{MMM}} & = & \int d^3 x \left[ \frac{1}{4} F_{L i j}^a F_{L i j}^a + \frac{1}{4}
  F_{R i j}^a F_{R i j}^a + (D_i \chi_L)^{\dag} (D_i \chi_L) + (D_{\mu}
  \chi_R)^{\dag} (D^{\mu} \chi_R) + V (\chi_L, \chi_R) \right],  \label{EF0}
\end{eqnarray}
where
\begin{eqnarray}
  F_{L i j}^a & = & \partial_i W_{L j}^a - \partial_j W_{L i}^a + g
  \epsilon^{a b c} W_{L i}^b W_{L j}^c, \\
  F_{R i j}^a & = & \partial_i W_{R j}^a - \partial_j W_{R i}^a + g
  \epsilon^{a b c} W_{R i}^b W_{R j}^c, \\
  D_{\mu} \chi_L & = & - \partial_i \chi_L - \frac{1}{2} i g \sigma^a W_{L
  i}^a \chi_L, \\
  D_{\mu} \chi_R & = & - \partial_i \chi_R - \frac{1}{2} i g \sigma^a W_{R
  i}^a \chi_R ,
\end{eqnarray}
with $W$ is the weak gauge boson, $g$ is the electroweak gauge coupling and $\sigma$ is the Pauli matrices.

From the energy functional in Eq.\eqref{EF0}, using the usual Euler-Lagrange method, we can write down the equations of motion for the fields as follows,
\begin{eqnarray}
  D_i (D_i \chi_L) & = & 4 \lambda \left( \chi_L^{\dag} \chi_L - \frac{1}{2}
  v^2_L \right) \chi_L + 2 \alpha \chi_L | \chi_R |^2, \label{DDchiL} \\
  D_i (D_i \chi_R) & = & 4 \lambda \left( \chi_R^{\dag} \chi_R - \frac{1}{2}
  v^2_R \right) \chi_R + 2 \alpha \chi_R | \chi_L |^2, \label{DDchiR} \\
  \partial_j F_{L i j}^a - g \epsilon^{a b c} W_{L j}^b F_{L i j}^c & = &
  \frac{1}{2} i g [\chi_L^{\dag} \sigma^a (D_i \chi_L) - (D_i \chi_L)^{\dag}
  \sigma^a \chi_L], \label{eomchiL}\\
  \partial_j F_{R i j}^a - g \epsilon^{a b c} W_{R j}^b F_{R i j}^c & = &
  \frac{1}{2} i g [\chi_R^{\dag} \sigma^a (D_i \chi_R) - (D_i \chi_R)^{\dag}
  \sigma^a \chi_R]. 
\end{eqnarray}
There are two scenarios related to the sphaleron solutions in this modified mirror model, depending on the value of $\alpha$, i.e., $\alpha \simeq 0$ and $\alpha \neq 0$. Below we will analyze each case.  

\subsection{The case $\alpha \simeq 0$}
In the case of $\alpha=0$ or very tiny, both SM and mirror sectors will completely decouple from each other. Thus, we can use the anzats solutions for the gauge boson and the Higgs fields as in the one-doublet Higgs model \cite{Klinkhamer:1984di} for each sector. In this case, their anzats solutions are written as,
\begin{subequations}\label{anz}
\begin{eqnarray}
  W_{L i}^a \sigma^a d x^i & = & - \frac{2 i}{g} f_L (g v_L r) d U (U)^{- 1}, \label{anz01}
  \\
  W_{R i}^a \sigma^a d x^i & = & - \frac{2 i}{g} f_R (g v_R r) d U (U)^{- 1}, \label{anz02}
  \\
  \chi_L & = & \frac{v_L}{\sqrt{2}} {h_L}  (g v_L r) U \left(\begin{array}{c}
    1\\
    0
  \end{array}\right), \label{anz03} \\
  \chi_R & = & \frac{v_R}{\sqrt{2}} {h_R}  (g v_R r) U \left(\begin{array}{c}
    1\\
    0
  \end{array}\right), \label{anz04}
\end{eqnarray}
\end{subequations}  
where $U$ is a unitary $2\times 2$ matrix given by,
\begin{eqnarray}
  U & = & \frac{1}{r} \left(\begin{array}{cc}
    z & x + i y\\
    - x + i y & z
  \end{array}\right) . 
\end{eqnarray}
In this scenario, each sector has its sphaleron solutions with each energy given by, 
\begin{eqnarray}
  E_i & = & \frac{4 \pi v_i}{g} \int_0^{\infty} \left[ 4 \left( \frac{d f_i}{d
  \xi_i} \right)^2 + \frac{8}{\xi^2_i} f_i^2 (1 - f_i)^2 + \frac{\xi^2_i}{2}
  \frac{d h_i^{\ast}}{d \xi_i} \frac{d h_i}{d \xi_i} + h_i^{\ast} h_i (1 -
  f_i)^2 + \frac{1}{4} \left( \frac{\lambda}{g^2} \right) \xi^2_i (h_i^{\ast}
  h_i - 1)^2 \right] d \xi_i , \label{Ei}
\end{eqnarray}
where $i=L,R$, and $\xi_i=g v_{i} r$ is a radial distance. In the above equation, $f_i$ and $h_i$ are the radial functions written as a function of the radial distance $\xi_i$ and the use of $\xi_i$ will simplify the expression for physical quantities that we shall consider later. The $B$ violating process in each sector is in thermal equilibrium if its rate is larger than the Hubble parameter. The $B$ violating process rate per unit volume due to thermal fluctuation is given by \cite{Arnold:1987mh,Dine:1991ck},
\begin{equation}
\gamma_{\text{sph}} =\left\{
 \begin{aligned}
    &\kappa' \alpha_{W}^{5} T_{i}^{4}, 
     \quad \text{for}\; T_{i}\geqslant T_{W_{i}}, \\        
    &2.8\times 10^{5} \kappa T_{i}^4 \left(\frac{\alpha_{W}}{4\pi}\right)^{4} \left[\frac{E_{\text{sph}}(T_{i})}{T_{i}\cal{B} }\right]^7 \exp\left(-\frac{E_{\text{sph}}(T_{i})}{T_{i}}\right),\quad \text{for}\; T_{i}< T_{W_{i}},
  \end{aligned}
  \right.
\end{equation}
where $\kappa'\simeq 25\pm 2$ is obtained from the lattice simulation \cite{Bodeker:1999zt}, $\kappa$ is the efficiency factor with a range $10^{-4}\lesssim \kappa \lesssim 10^{-1}$, $\alpha_{W}=g/4 \pi^2$ is the weak coupling constant and $\cal{B}$ is a varying function with a range value between $1.52$ until $2.72$ (the resulting integral over $\xi_{i}$ in Eq.\eqref{Ei}). $E_{\text{sph}}(T_{i})$ is the finite temperature sphaleron energy in each sector, $T_{i}$ is the temperature in each sector and $T_{W_{i}}$ is the electroweak phase transition temperature in each sector. If at high temperature the two sectors are in thermal equilibrium (before the inflation era), the $B$ violating process at the two sectors starts at the same temperature. But the $B$ violating process ends at different temperatures in each sector depending on the value of their vacuum expectation values or the electroweak phase transition temperature. The total $B$-violating process that could happen in the mirror sector is given by,
\begin{eqnarray}
    \Delta n &=& \int \gamma_{\text{sph}} dt = \int_{v_{i}}^{E_{\text{sph}}(v_{i})} \kappa' \alpha_{W}^{5} T^4  \frac{dT}{T^3} \frac{M_{Pl}}{1.66 \sqrt{g^{\ast}}}\nonumber \\ &\simeq& \left(\frac{4\pi v_{i} \cal{B} }{g} \right)^2 \frac{M_{Pl}}{1.66 \sqrt{g^{\ast}}} .
\end{eqnarray}
Thus, it is possible that in the mirror sector for very large $v_{R}$, the conversion $\Delta L\rightarrow \Delta B$ is not as effective as in the standard model sector. So the baryon asymmetry in the mirror sector is smaller than in the standard model sector. 

\subsection{The case $\alpha \neq 0$}
For this case, we only consider when $v_{R}\gg v_{L}$. In such a case when $\chi_{R}$ acquires its vacuum expectation value, the $\langle \chi_L \rangle \simeq 0$ and can be practically neglected. It is the same as in the one-doublet Higgs model \cite{Klinkhamer:1984di}. Then, the sphaleron energy in the mirror sector is given by, 
\begin{eqnarray}
  E_R & = & \frac{4 \pi v_R}{g} \mathcal{B}. 
\end{eqnarray}
On the other hand, when the $\chi_{L}$ acquires its VEV at a lower temperature, $\chi_{R}\simeq v_{R}$ and the equation of motion of $\chi_{L}$ in Eq.\eqref{DDchiL} becomes, 
\begin{eqnarray}
  D_i  (D_i \chi_L) & = & 4 \lambda \left( | \chi_L |^2 - \frac{1}{2} v^2_L +
  \frac{\alpha}{2\lambda} v_R^2 \right) \chi_L \nonumber\\
  & = & 4 \lambda \left( | \chi_L |^2 - \frac{1}{2} {v_L'}^2 \right) \chi_L,
  \\
  {v_L'}^2 & \equiv & v^2_L - \frac{ \alpha}{\lambda} v_R^2.
\end{eqnarray}
From Eq.\eqref{EF0}, we can compute the sphaleron energy in the standard model sector as follows,
\begin{equation}
  E_L  =  \int d^3 x \left[ \frac{1}{4} F_{L i j}^a F_{L i j}^a + (D_i
  \chi_L)^{\dag} (D_i \chi_L) + V (\chi_L, v_R) \right]  
 \end{equation}
 where $V (\chi_L, v_R)$ can be written as,
 \begin{eqnarray}
  V (\chi_L, v_R) & = & \lambda \left( | \chi_L |^2 - \frac{1}{2} v^2_L
  \right)^2 + \alpha | \chi_L |^2 v^2_R 
  \nonumber\\
  & = & \lambda \left[ | \chi_L |^4 - | \chi_L |^2 \left( v^2_L -
  \frac{\alpha}{\lambda} v^2_R \right) \right] = \lambda (| \chi_L |^4 - |
  \chi_L |^2 {v_L'}^2) \nonumber\\
  & = & \lambda \left( | \chi_L |^2 - \frac{1}{2} {v_L'}^2 \right)^2 
  .
\end{eqnarray}
Since the potential is similar like the case in Eq.\eqref{pot1} with $\alpha\simeq 0$ and neglecting $\chi_{R}=v_{R}$, but with shifted VEV $v_L'$, thus we can use the following new anzats solutions for the fields in the SM sector,
\begin{eqnarray}
  \chi_L & = & \frac{v_L'}{\sqrt{2}} {h_L}  (g v_L' r) U
  \left(\begin{array}{c}
    1\\
    0
  \end{array}\right), \\
  W_{L i}^a \sigma^a d x^i & = & - \frac{2 i}{g} f_L (g v_L' r) d U (U)^{- 1}.
\end{eqnarray}
Now the sphaleron energy in the SM sector is given by the following,
\begin{eqnarray}
  E_L & = & \frac{4 \pi v_L'}{g} \int_0^{\infty} \left[ 4 \left( \frac{d
  f_L}{d \xi_L'} \right)^2 + \frac{8}{{\xi_L'}^2} f_L^2 (1 - f_L)^2 +
  \frac{{\xi_L'}^2}{2} \frac{d h_L^{\ast}}{d \xi_L'} \frac{d h_L}{d \xi_L'} +
  h_L^{\ast} h_L (1 - f_L)^2 + \frac{1}{4} \left( \frac{\lambda}{g^2} \right)
  {\xi_L'}^2 (h_L^{\ast} h_L - 1)^2 \right] d \xi_L'  \nonumber \\
 &=&  \frac{4 \pi v_{L}' }{g} \mathcal{B} .
\end{eqnarray}

In the case that $\alpha\neq 0$ and $v_{R}\gg v_{L}$, the scenario is similar to the $\alpha=0$ where the $B$ violating process in the mirror sector terminates before the same process in the SM sector. The sphaleron energy in the SM sector is less than the case when there is no $\chi_{R}$ if $\alpha>0$. While the sphaleron energy in the SM sector is greater than the case when there is no $\chi_{R}$ if the coupling $\alpha<0$. For the case of $v_{R}\simeq v_{L}$, the sphaleron energy must be solved numerically.

\section{Conclusion}
In this paper, we have investigated the sphaleron solution in the modified mirror model. We focused only on the case where the doublet scalar of the mirror sector has a much larger VEV compared to the VEV of the SM Higgs, i.e., $v_{R} \gg v_{L}$. We investigated two scenarios related to the sphaleron solutions in this modified mirror model depending on the value of the coupling $\alpha$. In the case of $\alpha=0$, each sector has its sphaleron and it depends on the VEV of scalars for the each sector. In the case $\alpha \neq 0$, if $\alpha$ is positive, the sphaleron energy in the SM sector will decrease, while if $\alpha$ is negative the sphaleron energy will increase.

\begin{acknowledgments}
The work of A.S.A. is supported by Science and Technology Research Grant (STRG), Indonesian Torray Science Foundation (ITSF) 2022. 
\end{acknowledgments}

\providecommand{\noopsort}[1]{}\providecommand{\singleletter}[1]{#1}%

\end{document}